\documentclass[11pt,a4paper]{article}

\usepackage{authblk}
\usepackage[dvipsnames]{xcolor}
\usepackage[margin=1in,bottom=1in]{geometry}

\usepackage{amsmath}
\usepackage{amsthm}
\usepackage{amssymb}
\usepackage{amsfonts}
\usepackage{mathtools}
\usepackage{xparse}
\allowdisplaybreaks[1]

\usepackage{enumerate}
\usepackage{enumitem}
\usepackage{needspace}
\usepackage{float}
\usepackage[title]{appendix}

\usepackage{adjustbox}
\usepackage[nodisplayskipstretch]{setspace}

\usepackage[most]{tcolorbox}

\usepackage{indentfirst}
\usepackage{comment}

\usepackage{booktabs}
\usepackage{array}
\usepackage{colortbl}
\usepackage{subfigure}
\usepackage{makecell}

\usepackage[english]{babel}
\usepackage[T1]{fontenc}
\usepackage{xspace,microtype}

\usepackage[algoruled]{algorithm2e}
\SetAlgorithmName{Protocol}{Protocol}{List of Protocols}

\usepackage{tikz}
\usetikzlibrary{quantikz2}

\usepackage{xurl} 
\usepackage[pagebackref,colorlinks,linkcolor=NavyBlue,anchorcolor=blue,citecolor=NavyBlue]{hyperref}

\hypersetup{breaklinks=true}

\usepackage[capitalize,noabbrev]{cleveref}
\crefname{ineq}{inequality}{inequalities}
\creflabelformat{ineq}{#2{\upshape(#1)}#3}
\crefname{fact}{fact}{facts}
\crefname{equation}{equation}{equations}
\crefname{algocf}{Protocol}{Protocols}
\Crefname{algocf}{Protocol}{Protocols}
\crefname{remark}{remark}{remarks}
\crefname{conjecture}{conjecture}{conjectures}
\crefname{problem}{problem}{problems}
\numberwithin{equation}{section}

\usepackage{thmtools}
\makeatletter
\@ifundefined{newcounteralias}{}{%
  \renewcommand\thmt@autorefsetup{%
    \@xa\def\csname\thmt@envname autorefname
    \@xa\endcsname\@xa{\thmt@thmname}%
  }%
}
\makeatother

\usepackage{thm-restate}
\declaretheorem[style=plain,numberwithin=section]{theorem}

\declaretheorem[style=plain,numberlike=theorem]{lemma}
\declaretheorem[style=plain,numberlike=theorem]{corollary}
\declaretheorem[style=remark,numberlike=theorem]{remark}
\declaretheorem[style=plain,numberlike=theorem]{definition}
\declaretheorem[style=plain,numberlike=theorem]{proposition}

\declaretheorem[style=definition,numberlike=theorem]{problem,conjecture}
\numberwithin{equation}{section}

\makeatletter
\def\@buildmath#1{%
  \expandafter\def\csname bb#1\endcsname{\ensuremath{\mathbb{#1}}}%
  \expandafter\def\csname bf#1\endcsname{\ensuremath{\mathbf{#1}}}%
  \expandafter\def\csname sf#1\endcsname{\ensuremath{\mathsf{#1}}}%
  \expandafter\def\csname cal#1\endcsname{\ensuremath{\mathcal{#1}}}%
  \expandafter\def\csname rm#1\endcsname{\ensuremath{\mathrm{#1}}}%
  \expandafter\def\csname tt#1\endcsname{\ensuremath{\mathtt{#1}}}%
}
\def\@buildmathletters#1{%
  \ifx#1\relax\else
    \@buildmath{#1}%
    \expandafter\@buildmathletters
  \fi
} 
\@buildmathletters ABCDEFGHIJKLMNOPQRSTUVWXYZabcdefghijklmnopqrstuvwxyz\relax
\makeatother

\newcommand{\PSPACE}{\textnormal{\textsf{PSPACE}}\xspace}
\newcommand{\IP}{\textnormal{\textsf{IP}}\xspace}
\newcommand{\QIP}{\textnormal{\textsf{QIP}}\xspace}
\newcommand{\QIPtwo}{\textnormal{\textsf{QIP}(2)}\xspace}

\newcommand{\qqQAM}{\textnormal{\textrm{qq}\text{-}\textsf{QAM}}\xspace}
\newcommand{\qqStoqAM}{\textnormal{\textrm{qq}\text{-}\textsf{StoqAM}}\xspace}

\newcommand{\QAM}{\textnormal{\textsf{QAM}}\xspace}
\newcommand{\QCAM}{\textnormal{\textsf{QCAM}}\xspace}
\newcommand{\StoqMA}{\textnormal{\textsf{StoqMA}}\xspace}
\newcommand{\StoqAM}{\textnormal{\textsf{StoqAM}}\xspace}

\newcommand{\MA}{\textnormal{\textsf{MA}}\xspace}
\newcommand{\AM}{\textnormal{\textsf{AM}}\xspace}

\newcommand{\PP}{\textnormal{\textsf{PP}}\xspace}
\newcommand{\CH}{\textnormal{\textsf{CH}}\xspace}

\newcommand{\Tr}{\mathrm{Tr}}

\newcommand{\binset}{\{0,1\}}
\newcommand{\poly} {\operatorname{poly}}
\newcommand{\In}{\mathrm{in}}
\newcommand{\Out}{\mathrm{out}}

\DeclarePairedDelimiter\rbra{\lparen}{\rparen}
\DeclarePairedDelimiter\sbra{\lbrack}{\rbrack}
\DeclarePairedDelimiter\cbra{\{}{\}}
\DeclarePairedDelimiter\abs{\lvert}{\rvert}
\DeclarePairedDelimiter\norm{\lVert}{\rVert}

\renewcommand{\bra}[1]{\langle #1 \rvert}
\renewcommand{\ket}[1]{\lvert #1 \rangle}
\renewcommand{\braket}[3]{\langle #1 | #2 | #3 \rangle}
\newcommand{\ketbra}[2]{\ensuremath{\ket{#1}\!\bra{#2}}}
\newcommand{\innerprod}[2]{\langle #1 | #2 \rangle}

\RenewDocumentCommand{\Pr}{m g}{%
    \IfNoValueTF{#2}
        {\operatorname*{Pr}\sbra*{#1}}
        {\operatorname*{Pr}_{#1}\sbra*{#2}}%
}

\newcommand{\TQBF}{\textnormal{\textsc{TQBF}}\xspace}
\newcommand{\yes}{{\rm yes}}
\newcommand{\no}{{\rm no}}
\newcommand{\TV}{\mathrm{TV}}

\newcommand{\protocol}[2]{{#1}\!\rightleftharpoons\!{#2}}
\newcommand{\CIUniform}{\textnormal{\textsc{CIUniform}}\xspace}

\newcommand{\Now}{\mathrm{now}}
\newcommand{\Next}{\mathrm{next}}

\newcommand{\parheading}[1]{%
  \par\addvspace{1em}%
  \noindent\emph{#1}\enspace\ignorespaces%
}

\newtcolorbox{takeawaybox}{
    colback=purple!2,
    colframe=purple!50!black,
    boxrule=0.3pt,
    arc=0pt,
    left=60pt,
    right=60pt,
    top=6pt,
    bottom=6pt,
    before skip=8pt,
    after skip=8pt
}
\newcommand{\takeaway}[1]{%
\begin{takeawaybox}
\centering\itshape #1
\end{takeawaybox}%
}

\begin{document}

\setlength{\abovedisplayskip}{6pt}
\setlength{\belowdisplayskip}{6pt}

\title{Two messages suffice for quantum interactive proof systems}
\author[1]{Yupan Liu\thanks{Email: \url{yupan.liu@epfl.ch}}}
\affil[1]{%
  School of Computer and Communication Sciences,\protect\\
  \'Ecole Polytechnique F\'ed\'erale de Lausanne%
}

\date{}
\maketitle

\begin{abstract}
Quantum interactive proof systems can be parallelized to \emph{three} messages, as shown by \hyperlink{cite.KW00}{Kitaev and Watrous~(STOC 2000)}, perhaps their most striking distinction from classical counterparts. In this work, we establish that any quantum interactive proof system can further be parallelized to \emph{two} messages, with the verifier sending halves of EPR pairs (``quantum public coins'') and, after the prover's response, using only \emph{stoquastic} verification. Technically, we strengthen ${\sf QIP}(3) = {\sf QIP} = {\sf PSPACE}$ of \hyperlink{cite.JJUW11}{Jain, Ji, Upadhyay, and Watrous (JACM, 2011)} and fully characterize two-message quantum interactive proof systems by proving
\[ {\rm qq}\text{-}{\sf StoqAM} = {\rm qq}\text{-}{\sf QAM} = {\sf QIP}(2) = {\sf QIP} = {\sf PSPACE}. \]

Our main technical contributions show that \textsc{Close Image to Uniform} (\textsc{CIUniform}), which is naturally in the \emph{stoquastic} version of ${\rm qq}\text{-}{\sf QAM}$, is $\sf PSPACE$-hard via transforming the sum-check protocol of \hyperlink{cite.Shamir92}{Shamir~(JACM, 1992)} and \hyperlink{cite.Shen92}{Shen~(JACM, 1992)} into a \textsc{CIUniform} instance consisting of an efficient sampler whose image distribution is exactly \emph{uniform} for some input distribution when the underlying quantified Boolean formula is true. 
\end{abstract}

\section{Introduction}

The study of the quantum counterpart of \emph{interactive proof systems}~\cite{Babai85,GMR85} began in~\cite{Wat03}. In a quantum interactive proof system for a promise problem $(\calI_\yes,\calI_\no)$, a quantum polynomial-time \emph{verifier} interacts with an untrusted \emph{prover} whose computational power is unbounded, with at most polynomially many messages exchanged between them. Given an input $x\in\calI_\yes\cup\calI_\no$, the prover claims that $x\in\calI_\yes$, after which the two parties engage in an interactive protocol that ends with the verifier deciding whether to ``accept'' or ``reject'' the claim. The protocol has completeness $c$, meaning that there exists a prover strategy such that the verifier accepts with probability at least $c$ for \emph{yes} instances $x\in\calI_\yes$. The protocol also has soundness $s$, meaning that the verifier accepts with probability at most $s$ for \emph{no} instances, regardless of how the prover behaves. 

The celebrated $\QIP=\PSPACE$~\cite{JJUW11} and $\IP=\PSPACE$~\cite{LFKN92,Shamir92} theorems establish that quantum interactive proof systems have the same computational power as their classical counterparts when the completeness--soundness gap $c-s$ is at least inverse-polynomial. Still, quantum interactive proof systems exhibit a striking property called \emph{parallelization}: any quantum interactive proof system can be parallelized to use merely \emph{three} messages~\cite{KW00}, and the verifier's only message can be a single public coin~\cite{MW05}. In contrast, parallelizing classical interactive proof systems with \emph{polynomially} many messages to a constant number of messages would collapse the polynomial-time hierarchy to the second level~\cite{BHZ87}, since all \emph{constant}-message classical interactive proof systems can be parallelized to \emph{two} messages~\cite{Babai85,GS86}. 

\subsection{Main results}

In this work, we improve \emph{three}-message quantum interactive proof systems for \PSPACE{}~\cite{Wat03} to \emph{two} messages, and answer open problems in~\cite[Section 4]{Wat03} and~\cite[Section~4.5]{VW16}:

\begin{theorem}
    \label{thm:main-informal}
    $\qqStoqAM_1=\qqQAM=\QIPtwo=\QIP=\PSPACE.$
\end{theorem}

Here, the subscript $1$ denotes perfect completeness. Notably, perfect completeness for both \QIPtwo{} and \qqQAM{} was proved only very recently in~\cite{LV26}. Furthermore, combining our inclusion $\PSPACE\subseteq \QIPtwo$ with the three-message parallelization $\QIP\subseteq\QIP(3)$~\cite{KW00,KKMV09} and the containment $\QIP(3) \subseteq \PSPACE$~\cite{MW05,JJUW11} yields  
\takeaway{All quantum interactive proof systems with \emph{polynomially many} messages can be parallelized to \emph{two} messages.}

\paragraph{Implications on quantum public coins.}
The verifier's message in the two-message proof system underlying \Cref{thm:main-informal} consists merely of halves of EPR pairs, referred to as \emph{quantum public coins}, and the corresponding complexity class \qqQAM{} was introduced in~\cite{KLGN19}. Consequently, \Cref{thm:main-informal} improves the collapse theorem on generalized Arthur--Merlin proof systems~\cite[Theorem 1.7]{KLGN19}, in which the verifier's message contains either halves of EPR pairs (``$\rmq$'') or random bits (``$\rmc$''), reducing the \emph{four} categories to \emph{three}:
\begin{corollary}[Trichotomy theorem on generalized Arthur--Merlin games]
    \label{corr:trichotomy-thm}
    For any constant $m\geq 2$, the following trichotomy holds for $m$-message generalized quantum Arthur–Merlin games: 
    \begin{enumerate}[label={\upshape(\arabic*)}]
        \item \label{thmitem:tri-qqQAM} For any choice of message types $t_1, \cdots, t_m$ in $\cbra{\rmc,\rmq}$, if there exists an index $j\geq 2$ such that $t_j=\rmq$, then $t_m\cdots t_1\text{-}\QAM(m)=\PSPACE$. 
        \item \label{thmitem:tri-QAM} $\rmc\cdots \rmc\rmq\text{-}\QAM(m)=\rmc\rmq\text{-}\QAM \eqqcolon\QAM$. 
        \item \label{thmitem:tri-QCAM} $\rmc\cdots \rmc\text{-}\QAM(m)=\rmc\rmc\text{-}\QAM \eqqcolon\QCAM$. 
    \end{enumerate}
\end{corollary}
Since $\QAM \subseteq \mathsf{BP}\cdot\PP$~\cite[Theorem 4.7]{MW05}, proving that $\QAM=\qqQAM$ would collapse the counting hierarchy (\CH{}) to the second level. Thus, \Cref{corr:trichotomy-thm} indicates that
\takeaway{Unless \CH{} collapses, quantum public coins are \emph{strictly more powerful} than classical public coins in two-message quantum proof systems.}

\paragraph{Implications on stoquastic interactive proof systems.}
The verifier in the two-message proof system underlying \Cref{thm:main-informal} uses only \emph{stoquastic verification}, allowing only $\ket{0}$ and $\ket{+}$ ancillary qubits, classical reversible circuits, and a final Hadamard-basis measurement. When the verifier's message restricted to classical public coins, in the model denoted by \StoqAM{}, one can see that $\StoqAM=\AM$ by applying the averaging argument of~\cite[Section~5]{BBT06} to the verifiers obtained by fixing the public challenge. Since the one-message setting \StoqMA{} is contained in \AM{}~\cite{BBT06} and is believed to collapse to \MA{}~\cite{Liu21,AGL20}, \Cref{thm:main-informal} also reveals a sharp \emph{dichotomy} in the number of messages in stoquastic interactive proof systems: 
\begin{itemize}
    \item One message: \StoqMA{} appears not much more powerful than its classical counterpart \MA{}; 
    \item Two or more messages: \qqStoqAM{} is as powerful as its fully quantum counterpart \qqQAM{}, and thus captures the full power of classical interactive proof systems. 
\end{itemize}

\subsection{Proof overview}
To establish \Cref{thm:main-informal}, in particular $\PSPACE\subseteq \qqStoqAM_1$, we introduce \textsc{Close Image to Uniform} (\CIUniform{}, see \Cref{def:CIUniform}), a special case of \textsc{Close Image to Totally Mixed} introduced in~\cite{KLGN19}, as a bridge between \PSPACE{} and $\qqStoqAM_1$. An instance of \CIUniform{} consists of a polynomial-size sampler $\calS$, which takes $(u,r)\in\binset^{m_\In}\times\binset^{m_+}$ as input, with $r\sim U_{m_+}$ begin a uniformly random seed, and outputs $u'\in\binset^{m_\Out}$. The corresponding transition matrix is given by $M_\calS(u',u) = 2^{-m_+} \abs*{\cbra{r \colon \calS(u,r)=u'}}$. $\CIUniform[\eta]$ asks to distinguish between the following cases:
\begin{enumerate}
    \item[\textup{(yes)}] Some input distribution $p$ makes the given sampler's output $M_\calS p$ exactly \emph{uniform};
    \item[\textup{(no)}] Every input distribution $p$ makes $M_\calS p$ $\eta$-far from the uniform distribution $U_{m_\Out}$ with respect to the total variation distance. 
\end{enumerate}

To this end, we need to prove that (i) \CIUniform{} is \PSPACE{}-hard via the sum-check protocol used in~\cite{Shamir92,Shen92}; and (ii) \CIUniform{} is in $\qqStoqAM_1$. 

\paragraph{A recap of the sum-check protocol for \TQBF{}.}
We begin by considering \textsc{True Quantified Boolean Formula} (\TQBF{}), which is \PSPACE{}-complete~\cite[Theorem~4.3]{SM73}. In a quantified Boolean formula $\Phi=Q_1x_1\cdots Q_m x_m\,\phi(x_1,\ldots,x_m)$, each quantifier $Q_i\in\cbra{\exists,\forall}$ quantifies $x_i$ over $\binset$, and $\phi$ is a quantifier-free Boolean formula. 

As in~\cite[Section~3.1]{Wat03}, we use the sum-check protocol for \TQBF{}~\cite{Shen92}, which uses a sequence of polynomials $P_0,\ldots,P_T$ in $m'=O(|\Phi|)$ variables over a finite field $\bbF_q$, where $T$ denotes the number of rounds of the protocol. At the endpoints, the constant $P_0$ is $1$ if and only if $\Phi$ is true, while the verifier can evaluate $P_T$ directly. In any intermediate round $t<T$, the verifier does the following, where the operation in each round is either universal quantification, existential quantification, or \emph{degree reduction}:
\begin{enumerate}[label={\upshape(\arabic*)}]
    \setlength{\itemsep}{0pt}
    \item Receives the coefficients of a univariate polynomial $g$ of degree at most two,\footnote{In our construction, the multilinearity of $P_T$ and the degree reductions ensure that this bound holds in every round; see \Cref{lemma:efficient-TQBF-arithmetization}. A related individual-degree bound appears in~\cite[Lemma~9.2(a)]{BF91}.} claimed to be the univariate restriction $g_\star(X) = P_{t+1}(\alpha_1,\ldots,\alpha_{j_t-1},X,\alpha_{j_t+1},\ldots,\alpha_{m'})$ of $P_{t+1}$, where the operation in this round acts on coordinate $j_t$ and $\alpha$ denotes an assignment of $P_{t+1}$. 
    \item Uses $g(0)$ and $g(1)$ to check the current claim $P_t(\alpha)=\nu$ according to the round-$t$ operation.
    \item Evaluates $g$ at a field element $r\in\bbF_q$ chosen uniformly at random if the check passes, yielding the next-round claim $P_{t+1}(\alpha_1,\ldots,\alpha_{j_t-1},r,\alpha_{j_t+1},\ldots,\alpha_{m'})=g(r)$. 
\end{enumerate}
Importantly, degree reduction, as in~\cite{Shen92}, ensures the same degree bound for the true restriction $g_\star(X)$ in round $t$. If the current claim is false and the check passes, then $g-g_\star$ is a nonzero polynomial of degree at most two and thus $g(r)=g_\star(r)$ for at most two values of $r\in\bbF_q$. Consequently, the next-round claim is correct with probability at most $2/q$. Therefore, the soundness bound for the sum-check protocol in~\cite{Shen92} follows from a union bound over all $T$ rounds and holds for sufficiently large polynomially bounded $q$. 

\paragraph{\CIUniform{} is \PSPACE-hard via sum-check for \TQBF{}.} 
We reduce \TQBF{} to \CIUniform{}, and our proof construction now diverges from that of~\cite{Wat03}. In the three-message quantum proof system of~\cite[Section~3.2]{Wat03}, the verifier first receives complete sum-check transcripts in superposition, and then checks the \emph{local consistency} of successive sum-check claims within each transcript.
Instead, our construction checks the \emph{global consistency} across the sum-check rounds. 

To this end, we construct an efficient sampler $\calS$ based on a one-step sum-check transition, which takes a tuple $(t,\alpha,g)$ as input, consisting of the round index $t$, an evaluation point $\alpha\in\bbF_q^{m'}$, and the three coefficients of the claimed univariate polynomial $g$, and outputs a tuple $(t',\alpha',\nu)$ representing the claim $P_{t'}(\alpha')=\nu$. This global consistency gives the condition tested by our $\qqStoqAM_1$ proof system: there exists an input distribution that makes the sampler's output exactly uniform when the underlying quantified Boolean formula $\Phi$ is true.

More specifically, for \emph{yes} instances with $\Phi$ true, choose the pair $(t,\alpha)$ uniformly at random and take $g=g_\star$ for $t<T$ while $g=0$ for $t=T$. The sampler starts from a uniform tuple $(t,\alpha,\nu)$ and replaces it by the next-round claim only when $\nu=P_t(\alpha)$. If no replacement occurs, $(t',\alpha')=(t,\alpha)$. When replacement occurs, the next-round index $t'$ moves forward cyclically: 
\begin{enumerate}[label={\upshape(\roman*)}]
    \setlength{\itemsep}{0pt}
    \item If $t<T$, the next round is $t'=t+1$, and $\alpha'$ is obtained by replacing the coordinate $\alpha_{j_t}$ used in round $t$ with $r\in\bbF_q$ chosen uniformly at random;
    \item If $t=T$, the sampler returns to round $0$ and keeps $\alpha'=\alpha$.
\end{enumerate}
The pair $(t',\alpha')$ is therefore uniformly distributed. Since the replacement occurs with probability $1/q$ at each $(t,\alpha)$ and the next-round claim is correct also when the sampler returns to round $0$, the next-round claim has the same distribution as the claim being replaced. Thus, the sampler's output remains exactly uniform. 

\vspace{1em}
For \emph{no} instances with $\Phi$ false, we fix an arbitrary input distribution $p$ and bound the total variation distance $\TV(M_\calS p,U_{m_\Out})$ between the sampler's output distribution $M_\calS p$ and the uniform distribution $U_{m_\Out}$. To this end, we consider the following two events under $U_{m_\Out}$: 
\begin{enumerate}[label={\upshape(\roman*)}]
    \setlength{\itemsep}{0pt}
    \item \label{thmitem:TQBF-to-CIUniform-no-i}The event that round $T$ is chosen ($t'=T$) occurs with probability $1/(T+1)$ when $(t',\alpha',\nu)$ is drawn uniformly at random from $\cbra{0,\ldots,T}\times\bbF_q^{m'}\times\bbF_q$;
    \item \label{thmitem:TQBF-to-CIUniform-no-ii}The event that $P_{t'}(\alpha')=\nu$ occurs with probability $1/q$, for each fixed $t'$ and $\alpha'$. 
\end{enumerate}
Let $(t',\alpha',\nu)\sim M_{\calS}p$. Taking the corresponding weighted average of the differences between the probabilities of these events under $U_{m_\Out}$ and $M_{\calS} p$, each of which is at most $\TV(M_\calS p,U_{m_\Out})$, we obtain the following bound
\begin{equation}
    \label{eq:TQBF-to-CIUniform-no-intro}
    \TV(M_\calS p,U_{m_\Out}) \geq \frac{1}{q+1}\abs*{ \frac{1}{T+1} - \Pr{t'=T} } + \frac{q}{q+1} \abs*{\frac{1}{q} - \Pr{P_{t'}(\alpha')=\nu}}.
\end{equation}

Further analysis of our sampler construction yields lower bounds on $\abs*{1/(T+1)-\Pr{t'=T}}$ for event \ref{thmitem:TQBF-to-CIUniform-no-i} and on $\abs*{1/q - \Pr{P_{t'}(\alpha')=\nu}}$ for event \ref{thmitem:TQBF-to-CIUniform-no-ii}. Plugging these bounds into \Cref{eq:TQBF-to-CIUniform-no-intro}, we see that the unknown term $\Pr{t=T}$ in the resulting lower bounds cancels due to the weights in \Cref{eq:TQBF-to-CIUniform-no-intro}. Taking $q=8(T+1)$, we conclude that $\TV(M_\calS p,U_{m_\Out}) \geq 1/(4q(T+1))$ for every input distribution $p$.

The remaining challenge is to ensure that the resulting $\CIUniform[\eta]$ instance has an inverse-polynomial parameter $\eta=1/(32(T+1)^2)$, which is crucial for the resulting \qqStoqAM{} proof system to achieve a constant promise gap via parallel repetition.\footnote{For instance, for a field of exponential cardinality, as in~\cite[Theorem~6]{Shamir92}, $\TV(M_\calS p,U_{m_\Out})$ and thus $\eta$ would be exponentially small, so the resulting \qqStoqAM{} proof system would also have only an exponentially small promise gap.}
Using degree reduction in~\cite{Shen92}, one can take $T+1=O(|\Phi|^2)$ and hence $\eta=1/O(|\Phi|^4)=1/\poly(n)$, as desired. 

\paragraph{A \qqStoqAM{} proof system for \CIUniform{}.}
Given a $\CIUniform[\eta]$ instance, we have the corresponding sampler $\calS \colon \binset^{m_\In}\times\binset^{m_+} \rightarrow \binset^{m_\Out}$ and its reversible circuit implementation $U_\calS$ such that $U_\calS\ket{u,r,w} = \ket{u,r,w\oplus\calS(u,r)}$ for each $u\in\binset^{m_\In}$, $r\in\binset^{m_+}$, and $w\in\binset^{m_\Out}$. We consider the following $\qqStoqAM_1$ proof system: 
\begin{enumerate}[label={\upshape(\arabic*)}]
    \setlength{\itemsep}{0pt}
    \item Prepare $m_\Out$ EPR pairs across the registers $(\sfB,\sfB'')$ and send $\sfB''$ to the prover.
    \item Receive the $m_\In$-qubit input register $\sfA$ and the $m_+$-qubit seed register $\sfR$, and apply $U_\calS^\dagger$ to the registers $(\sfA,\sfR,\sfB)$.
    \item Apply the SWAP test, controlled by a single-qubit register $\sfO$ initialized to $\ket{+}$, between the registers $(\sfR,\sfB)$ and $(\sfR',\sfB')$, where $\sfR'$ and $\sfB'$ are initialized to $\ket{+}^{\otimes m_+}$ and $\ket{0}^{\otimes m_\Out}$, respectively. Accept if and only if the SWAP test accepts. 
\end{enumerate}

We then explain this proof system. For \emph{yes} instances, choose a distribution $p$ such that $M_\calS p=U_{m_\Out}$. Upon receiving $\sfB''$, an honest prover maps each basis state $\ket{u'}$ to a quantum state proportional to $\sum_{(u,r):\calS(u,r)=u'}\sqrt{p(u)}\ket{u}_{\sfA}\ket{r}_{\sfR}$ via an isometry. Since $(M_\calS p)(u')=2^{-m_\Out}$ for every $u'$, the normalization factor is identical for every $u'$. Consequently, after applying $U_\calS^\dagger$, the verifier obtains the state $\rbra[\big]{ \sum_u\sqrt{p(u)}\ket{u}_\sfA }\ket{+}_\sfR^{\otimes m_+}\ket{0}_\sfB^{\otimes m_\Out}$. Since the reduced density matrix on $(\sfR,\sfB)$ matches that on $(\sfR',\sfB')$, the SWAP test accepts with certainty. 

For \emph{no} instances, $\TV(M_\calS p,U_{m_\Out})\geq\eta$ for every input distribution $p$. Fix an arbitrary prover strategy and let $\ket{\psi}$ be the joint pure state on $(\sfA,\sfR,\sfB,\sfQ)$ just before the verifier applies $U_\calS^\dagger$, where $\sfQ$ denotes the prover's private register. Since $\sfB$ is private to the verifier and consists of halves of the EPR pairs, measuring $\sfB$ on $\ket{\psi}$ gives $U_{m_\Out}$. In contrast, consider any pure state $\ket{\xi}$ on $(\sfA,\sfR,\sfB,\sfQ)$ obtained by applying $U_\calS$ with $(\sfR,\sfB)$ initialized to $\ket{+}^{\otimes m_+}\ket{0}^{\otimes m_\Out}$, allowing $\sfA$ to be entangled with $\sfQ$. Measuring $\sfB$ on $\ket{\xi}$ gives $M_\calS p$ for some $p$. Since measurement in the computational basis cannot increase trace distance, it follows that
\[ \eta \leq \TV(M_\calS p, U_{m_\Out}) \leq \sqrt{1-\abs{\innerprod{\psi}{\xi}}^2}. \] 
Therefore, $\abs{\innerprod{\psi}{\xi}}^2 \leq 1-\eta^2$ holds for every such $\ket{\xi}$, and the SWAP test thus accepts with probability at most $(1+(1-\eta^2))/2=1-\eta^2/2$, as desired. 


\section{Preliminaries}
\label{sec:preliminaries}

We assume that the reader is familiar with quantum information and computation, and the textbook~\cite{NC10} serves as a good starting point. We recommend~\cite{VW16} for a self-contained survey of quantum interactive proof systems. 

Throughout the paper, all logarithms are base-two. 
We use $U_\ell$ to denote the uniform distribution on $\ell$ bits, and write $\bbI[E]$ for the indicator function that equals $1$ if and only if the event $E$ occurs. 
For any probability distributions $p_0$ and $p_1$ over a set $S$, the total variation distance between them is defined by $\TV(p_0,p_1) \coloneqq \frac{1}{2} \sum_{x\in S} \abs*{p_0(x) - p_1(x)}$. 

\subsection{Two-message quantum proof systems with quantum public coins}
A \textit{promise problem} $\calI = (\calI_{\yes}, \calI_{\no})$ satisfies that $\calI_{\yes} \cap \calI_{\no} =\emptyset$ and $\calI_{\yes} \cup \calI_{\no} \subseteq \binset^*$. For convenience, we write $x\in\calI$ for $x\in \calI_{\yes} \cup \calI_{\no}$. 
Given a promise problem $\calI$, we write $(\protocol{P}{V})(x)$ for a quantum interactive proof system on input $x\in \calI$. We then recap the definition of generalized Arthur--Merlin proof systems from \cite[Section 3]{KLGN19}:
\begin{definition}[\qqQAM{}]
\label{def:qqQAM}
Let $c(n)$ and $s(n)$ be functions of the input length $n \coloneqq |x|$ such that $0 \leq s(n) < c(n) \leq 1$. Let $\ell(n)$ be a positive integer-valued function such that $\ell(n) \leq \poly(n)$.
A promise problem $\calI = (\calI_{\yes}, \calI_{\no})$ is in $\qqQAM\sbra*{\ell,c,s}$, if there exists an efficient quantum verifier $V(x)=(V_1(x),V_2(x))$ and a polynomial $p$, where $V_1$ prepares $p(n)$ EPR pairs and sends their second halves to the prover, and stores their first halves in $\sfV$, and $V_2$ acts on the message register $\sfM$ and $\sfV$, such that:
\begin{itemize}
    \item \textbf{\emph{Completeness}}. For every $x \in \calI_{\yes}$, there exists a prover $P(x)$ such that, upon receiving the halves of the EPR pairs, $P(x)$ returns an $\ell(n)$-qubit state and
    \[\Pr{(\protocol{P}{V})(x)\text{ accepts}} \geq c(n).\]
    \item \textbf{\emph{Soundness}}. For any $x \in \calI_{\no}$ and any prover $P(x)$ that returns an $\ell(n)$-qubit state, 
    \[\Pr{(\protocol{P}{V})(x)\text{ accepts}} \leq s(n).\]
\end{itemize}
\end{definition}

Since \qqQAM{} achieves perfect completeness~\cite[Theorem~1.2]{LV26} and admits error reduction~\cite[Lemma~3.6]{KLGN19}, we adopt the convention 
\[ \qqQAM \coloneqq \cup_{\ell(n) \leq \poly(n)} \qqQAM[\ell,1,1/2]. \] 
We also omit the parameter $\ell$ and when convenient and write 
\[ \qqQAM[c,s] \coloneqq \cup_{\ell(n) \leq \poly(n)} \qqQAM[\ell,c,s]. \]

\subsection{Stoquastic algorithmic toolkit}

We need the following definition from~\cite[Definition~2.11]{LW26}: 
\begin{definition}[Stoquastic verification circuit, implicit in~{\cite[Definition~4]{BBT06}}]
    \label{def:stoquastic-verification}
    We say that a quantum circuit $Q$ is a \emph{stoquastic verification circuit} if $Q$ acts on the registers $(\sfW,\sfA_0,\sfA_+)$, uses only Toffoli, CNOT, and X gates before the final measurement, and then measures the designated output qubit in the Hadamard basis. Here, the workspace register $\sfW$ contains an $m_w$-qubit state, and the ancillary registers $\sfA_0$ and $\sfA_+$ are initialized to $\ket{0}^{\otimes m_0}$ and $\ket{+}^{\otimes m_+}$, respectively. The circuit $Q$ accepts if and only if the final measurement gives outcome $+$.
\end{definition}

The SWAP test was originally proposed for pure states in~\cite{BCWdW01} and was later shown to extend to mixed states~\cite{KMY09}: 
\begin{lemma}[SWAP test for mixed states, adapted from~{\cite[Proposition 9]{KMY09}}]
    \label{lemma:swap-test}
    Let $\rho_0$ and $\rho_1$ be two $m$-qubit quantum states, which may in general be \emph{mixed}. Then there exists a $(2m+1)$-qubit stoquastic circuit that accepts with probability $\rbra*{1+\Tr(\rho_0\rho_1)}/2$, and uses a single copy of each $\rho_0$ and $\rho_1$, together with $m$ Toffoli gates and $2m$ CNOT gates.
\end{lemma}


\section{\CIUniform{} is \PSPACE{}-hard}
\label{sec:TQBF-to-CIUniform}

We begin by defining the promise problem \textsc{Close Image to Uniform}: 

\begin{definition}[\textsc{Close Image to Uniform}, $\CIUniform$]\label{def:CIUniform}
Let $\calS$ be a polynomial-size Boolean circuit that takes an $m_{\In}$-bit string $u$ and an $m_+$-bit random seed $r$ as input and outputs an $m_\Out$-bit string $u'$. Equivalently, for each $u\in\binset^{m_\In}$, $\calS$ serves as a sampler that draws $r\sim U_{m_+}$ independently and outputs $\calS(u,r)$. The corresponding transition matrix and output distribution for an input distribution $p$ are given by 
\[ M_\calS(u',u)=2^{-m_+}\abs*{\cbra{r\colon\calS(u,r)=u'}}
    \quad\text{and}\quad
    (M_\calS p)(u')=\sum_{u\in \binset^{m_{\In}}} M_\calS(u',u)p(u). \]
Let $n\coloneqq m_{\In}+m_{\Out}$ be the input length, and let $\eta(n)$ be an efficiently computable function such that $0\leq \eta(n) \leq 1$. 
The promise problem $\CIUniform[\eta]$ asks to decide whether
\begin{enumerate}
    \item[\textup{(yes)}] There exists an input distribution $p$ such that $M_\calS p=U_{m_\Out}$.
    \item[\textup{(no)}] For all input distributions $p$, it holds that $\TV(M_\calS p,U_{m_\Out})\geq\eta$.
\end{enumerate}
\end{definition}

In this section, we reduce \TQBF{} to \CIUniform{} and thereby prove the \PSPACE{}-hardness:

\begin{theorem}[\CIUniform{} is \PSPACE{}-hard]
    \label{thm:CIUniform-PSPACE-hard}
    There exists some polynomial $p(n)$ such that
    \[ \CIUniform[1/p(n)] \text{ is } \PSPACE{}\text{-hard}. \] 
\end{theorem}

For the remainder of this section, we first recap the sum-check protocol for \TQBF{} in \Cref{subsec:sum-check-recap}, in the specific format required by our construction, and then construct a sampler from the one-step transitions of the protocol in \Cref{subsec:sum-check-transcript-sampler}.

\subsection{The sum-check protocol for \TQBF{}}
\label{subsec:sum-check-recap}

Let $\Phi=Q_1x_1\cdots Q_m x_m\,\phi(x_1,\ldots,x_m)$, with $m\geq 1$, be a \TQBF{} instance, where each $Q_i\in\cbra{\exists,\forall}$ quantifies $x_i$ over $\binset$ and $\phi$ is a quantifier-free formula over AND, OR, and NOT. We say that $\Phi$ is a \emph{yes} instance if and only if $\Phi$ is true. 

In the sum-check protocol for \TQBF{} in~\cite{Shen92}, the interactive verifier reduces a claim about whether $\Phi$ is true to a polynomial evaluation that it can directly check. Let $T$ be the number of rounds in this protocol. 
Applying the arithmetization on $\Phi$, the verifier constructs a Boolean circuit for a multilinear polynomial $P_T$ over $\bbF_q$ in $m'$ variables and a list of operations $\calO_0,\ldots,\calO_{T-1}$. Starting with $P_T$, define the intermediate polynomials backwards by
\begin{equation}
    \label{eq:polynomial-claims}
    \forall t\in\cbra{T-1,\ldots,0}, \quad P_t \coloneqq \mathcal{O}_t P_{t+1}. 
\end{equation}
Each operation $\mathcal{O}_t$ acts on one coordinate $j_t$. For an evaluation point $\alpha\in\bbF_q^{m'}$, we write $\alpha[j_t \leftarrow X]$ for the tuple obtained by replacing its $j_t$-th coordinate $\alpha_{j_t}$ by $X$. Each $\calO_t$ is one of three types: the universal quantification $A_{j_t}$, the existential quantification $E_{j_t}$, or the degree reduction $R_{j_t}$. 
For round $t$, define $L_{t,\alpha}\colon\bbF_q^2\to\bbF_q$ by
\[L_{t,\alpha}(x,y)=
 \begin{cases}
  xy & \calO_t=A_{j_t},\\
  x+y-xy & \calO_t=E_{j_t},\\
  (1-\alpha_{j_t})x+\alpha_{j_t} y & \calO_t=R_{j_t}.
 \end{cases}\]
For any polynomial $P$, the corresponding operator at round $t$ satisfies
\[ (\calO_tP)(\alpha) = L_{t,\alpha} \rbra*{ P(\alpha[j_t \leftarrow 0]),P(\alpha[j_t \leftarrow 1]) }. \]
We will show later in \Cref{lemma:efficient-TQBF-arithmetization} that $P_0$ equals to the truth value of $\Phi$, and every $P_t$ has individual degree at most two. 
We now recap the protocol for \TQBF{}~\cite{Shen92} in \Cref{protocol:sum-check}.

\LinesNotNumbered
\begin{algorithm}[ht!]
    \caption{The sum-check protocol for \TQBF{}.}
    \label[algocf]{protocol:sum-check}
    \SetEndCharOfAlgoLine{.}
    \SetAlgoVlined
    \setlength{\parskip}{5pt}
    \SetKwFor{For}{For}{:}{}
    \SetKwIF{If}{ElseIf}{Else}{If}{:}{ELSE IF}{Else:}{}
    \textbf{1.} Start at $\alpha=0^{m'}$ with the claim $P_0(\alpha)=1$\;
    \smallskip
    \For{\textup{round } $t \leftarrow 0$ \KwTo $T-1$}{
        Let $j_t$ denote the coordinate on which $\calO_t$ acts. Let $P_t(\alpha)=\nu$ be the  current claim\;
        \textbf{2.a} Receive the three coefficients of a univariate polynomial $g$ of degree at most two, claiming that $g$ equals the restriction $g_\star(X)=P_{t+1}(\alpha[j_t\leftarrow X])$\;
        \textbf{2.b} \eIf{$\nu\ne L_{t,\alpha}(g(0),g(1))$}{
            Reject\;
        }{
            Draw a fresh uniform $r\in\bbF_q$ and proceed with the claim $P_{t+1}(\alpha[j_t\leftarrow r])=g(r)$\;
        }
    }
    \smallskip
    \textbf{3.} In round $T$, accept exactly when the claimed value equals $P_T(\alpha)$. 
\end{algorithm}

\paragraph{Constructing the polynomials.}
We now construct the polynomials $P_T,P_{T-1},\ldots,P_0$ from $\Phi$. To make $P_T$ multilinear, we use choice bits to encode the evaluation of $\phi$ (\Cref{prop:formula-evaluation-game}) and apply interpolation to express each choice as a polynomial (\Cref{prop:one-bit-interpolation}).
In particular, we push negations down to the variables and write the quantifier-free formula $\phi$ as a tree. Each internal node is a binary AND or OR gate, with two children as inputs, and the leaves are literals $x_i$ or $\neg x_i$, or Boolean constants. 
For each internal gate $h$, we introduce a choice bit $z_h$: starting at the root, $z_h=0$ selects the left input and $z_h=1$ selects the right input, until a leaf is reached. 

\begin{proposition}[Evaluating a Boolean formula by quantified choices, adapted from~{\cite[Section~2]{Hintikka82}}]
    \label{prop:formula-evaluation-game}
    Fix a Boolean assignment $\alpha$ to $x_1,\ldots,x_m$. Consider the predicate that the leaf selected by the choice bits has value $1$ under $\alpha$. Assign quantifiers to the choice bits $z_h$ in parent-before-child order, using $\exists$ for each OR gate and $\forall$ for each AND gate. The resulting quantified predicate is true if and only if $\phi$ evaluates to $1$ under $\alpha$.
\end{proposition}

\begin{proposition}[Interpolation at zero and one, adapted from~{\cite[Equation~(4.15) and Remark~4.5]{BFL91}}]
    \label{prop:one-bit-interpolation}
    Let $P_L$ and $P_R$ be polynomials over $\bbF_q$, and let $z$ be a fresh variable. The unique polynomial affine in $z$ with values $P_L$ at $z=0$ and $P_R$ at $z=1$ is $(1-z)P_L+zP_R$. If $P_L$ and $P_R$ are multilinear, the resulting polynomial is also multilinear in all variables, even if the two polynomials share variables. 
\end{proposition}

To construct $P_t$ for each $t<T$ while keeping the degree bounded, we interleave quantifier elimination with degree reduction~\cite{Shen92}, so that every $P_t$ has individual degree at most two:

\begin{lemma}[Degree reduction for $\TQBF$, adapted from~{\cite[Section~2]{Shen92}}]
    \label{lemma:quantified-degree-reduction}
    Consider a quantified formula $Q_1x_1\cdots Q_{m'}x_{m'}[P=1]$, where $P$ is a multilinear polynomial over $\bbF_q$ taking Boolean values on $\binset^{m'}$. Starting with $P$, for $j = m',\ldots,1$ apply $R_1,\ldots,R_{j-1}$ and then apply $A_j$ if $Q_j=\forall$ or $E_j$ if $Q_j=\exists$.\footnote{We omit the $R_j$ used in~\cite{Shen92} immediately before applying $A_j$ or $E_j$, since $R_j$ preserves the two restrictions obtained by setting $x_j\in\binset$, on which both $A_j$ and $E_j$ depend.} Every intermediate polynomial has individual degree at most two. The final polynomial is a constant equal to the truth value of the formula.
\end{lemma}

With \Cref{prop:formula-evaluation-game,prop:one-bit-interpolation,lemma:quantified-degree-reduction} in hand, we obtain the following efficiently constructible arithmetization of a \TQBF{} instance $\Phi$:

\begin{lemma}[Efficient arithmetization of \TQBF{}]
    \label{lemma:efficient-TQBF-arithmetization}
    Given $\Phi$, one can construct in polynomial time a finite field $\bbF_q$, a Boolean circuit implementation for a multilinear polynomial $P_T$ over $\bbF_q$ in $m'=O(|\Phi|)$ variables, and a list of $T$ operations $\calO_t\in\cbra{A_{j_t},E_{j_t},R_{j_t}}$ indexed by $t\in\cbra{0,\ldots,T-1}$, where $T+1=O(|\Phi|^2)$ is a power of two and $q=8(T+1)$. The polynomials defined by \Cref{eq:polynomial-claims} have individual degree at most two, and $P_0=\bbI\sbra{\Phi\text{ is true}}$.
\end{lemma}

\begin{proof}
We begin by identifying all variables in $P_T$. To this end, we append the quantifiers associated with the choice bits $z_h$ to the prefix of $\Phi$ as in \Cref{prop:formula-evaluation-game}. There is one choice bit per internal gate of $\phi$, so the extended prefix has $m'=O(|\Phi|)$ variables. 

Next, we construct $P_T$ from the leaves of $\phi$ to the root, over any finite field. Represent the literals $x_i$ and $\neg x_i$ by $x_i$ and $1-x_i$, respectively, and each constant by its Boolean value. At each internal gate $h$ with child polynomials $P_L$ and $P_R$, use $(1-z_h)P_L+z_hP_R$. By \Cref{prop:one-bit-interpolation}, the polynomial at the root is multilinear and, on Boolean inputs, equals the value of the selected leaf, giving an arithmetic circuit for $P_T$ with $O(|\phi|)$ gates. Fix an arbitrary Boolean assignment $\alpha$ to the variables in $\Phi$ and write $z$ for the tuple of choice bits. By \Cref{prop:formula-evaluation-game}, applying the choice-bit quantifiers to $P_T(\alpha,z)=1$ gives the truth value of $\phi(\alpha)$. Applying the original quantifiers $Q_1x_1\cdots Q_mx_m$ therefore gives the truth value of $\Phi$.

We subsequently turn the extended prefix into round operations. Apply \Cref{lemma:quantified-degree-reduction} to $P_T$ with this extended prefix. Eliminating the variable in position $j$ of the extended prefix uses $j-1$ degree reductions and one quantifier operation, for a total of $\sum_{j=1}^{m'}j=m'(m'+1)/2$ operations. Let $T+1$ be the smallest power of two greater than $m'(m'+1)/2$. Hence, $T+1 \leq m'(m'+1) = O(|\Phi|^2)$. Append copies of $R_1$ until the list has $T$ operations. This padding preserves $\bbI\sbra{\Phi\text{ is true}}$, as $R_1$ fixes the final constant. For $k=1,\ldots,T$, let $\calO_{T-k}$ denote the $k$-th operation in the padded list. Then, \Cref{eq:polynomial-claims} gives $P_0=\bbI\sbra{\Phi\text{ is true}}$, and the fact that every $P_t$ has individual degree at most two follows from \Cref{lemma:quantified-degree-reduction}.

Finally, we implement the resulting arithmetic circuit for $P_T$ as a Boolean circuit. Set $q=8(T+1)$. We represent $\bbF_q$ as $\bbF_2[X]/(f)$, where $f$ is a monic irreducible polynomial of degree $\log{q}$. There are $q$ monic candidates for $f$ and fewer than $q$ monic polynomials of smaller positive degree to test as divisors. Thus, trial division finds such an $f$ in deterministic time polynomial in $q$, hence in $|\Phi|$. Represent each field element by the $\log{q}$ binary coefficients of its
remainder modulo $f$. In this representation, addition is bitwise XOR,
and multiplication is polynomial multiplication followed by reduction
modulo $f$. Therefore, each arithmetic gate in the circuit for $P_T$ can be
implemented with $\poly(\log{q})$ Boolean gates.
\end{proof}

\paragraph{Analysis of \Cref{protocol:sum-check}.}
Fix a round $t<T$ and an evaluation point $\alpha$. By \Cref{lemma:efficient-TQBF-arithmetization}, the restriction $g_\star$ in \Cref{protocol:sum-check} has degree at most two. The definition of $L_{t,\alpha}$ and \Cref{eq:polynomial-claims} give $L_{t,\alpha}(g_\star(0),g_\star(1))=(\calO_t P_{t+1})(\alpha)=P_t(\alpha)$. 
If $\Phi$ is true, the initial claim $P_0(0^{m'})=1$ is correct. Whenever the current claim is correct, the reply $g=g_\star$ passes the verifier's check and gives the correct next-round claim for every $r$, since $g_\star(r)=P_{t+1}(\alpha[j_t\leftarrow r])$. Consequently, the terminal claim is correct and the verifier accepts with certainty.

Now fix any false current claim $P_t(\alpha)=\nu$ and any reply $g$ that passes the verifier's check. Then, we obtain the inconsistency 
\[ L_{t,\alpha}(g(0),g(1))=\nu\ne P_t(\alpha)=L_{t,\alpha}(g_\star(0),g_\star(1)). \] 
Hence, $g-g_\star$ is a nonzero polynomial of degree at most two. The verifier draws $r\in\bbF_q$ uniformly at random after receiving $g$, so the probability that the next-round claim is correct is
\begin{equation}\label{eq:false-successor}
    \Pr{r\sim U_{\log{q}}}{g(r)=g_\star(r)}
    =\frac{\abs*{\cbra{r\in\bbF_q\colon (g-g_\star)(r)=0}}}{q}
    \leq\frac{2}{q}.
\end{equation}
If $\Phi$ is false, the initial claim is false, and acceptance requires a correct terminal claim. An accepting execution of the sum-check protocol must therefore contain a round in which a false current claim gives a correct next-round claim. By \Cref{eq:false-successor}, the probability of such a transition is at most $2/q$ in each round, even conditioned on the transcript through the prover's reply. A union bound over the $T$ rounds gives acceptance probability at most 
\[ 2T/q=T/(4(T+1))<1/4. \] 

\subsection{Constructing the sampler from one-step sum-check transitions}
\label{subsec:sum-check-transcript-sampler}

Fix a \TQBF{} instance $\Phi$ and let $\calS=\calS_\Phi$. An input $u \coloneqq (t,\alpha,g)$ specifies a round index $t\in\cbra{0,\ldots,T}$, an evaluation point $\alpha\in\bbF_q^{m'}$, and the three coefficients of a univariate polynomial $g$ of degree at most two. At $t=T$, the sampler ignores $g$. The sampler's output $u' \coloneqq (t',\alpha',\nu)$ represents the claim $P_{t'}(\alpha')=\nu$. For $r\in\bbF_q$ chosen uniformly at random, define the tuples representing the current claim $C_{\Now}(u)$ and the next-round claim $C_{\Next}(u,r)$ by
\begin{subequations}
\label{eq:fragment}
\begin{align}
    C_{\Now}(u)& \coloneqq \begin{cases}
    (t,\alpha,L_{t,\alpha}(g(0),g(1))), &t<T\\
    (T,\alpha,P_T(\alpha)), &t=T
    \end{cases},\\
    C_{\Next}(u,r)& \coloneqq \begin{cases}
    (t+1,\alpha[j_t\leftarrow r],g(r)), & t<T\\
    (0,\alpha,1), & t=T
    \end{cases}.
\end{align}
\end{subequations}
For $t<T$, the claimed value in $C_{\Now}(u)$ is exactly the one for which the reply $g$ passes the verifier's check in \Cref{protocol:sum-check}. At $t=T$, $C_{\Now}(u)$ uses the value $P_T(\alpha)$, and $C_{\Next}(u,r)$ returns to round $0$ with the claim $P_0(\alpha)=1$. In the wraparound case $t=T$ of \Cref{eq:fragment}, we keep the evaluation point $\alpha$ unchanged because $P_0$ is constant. 

Choose $v\in\bbF_q$ uniformly at random and form the tuple $(t,\alpha,v)$. Replace this tuple by the next-round claim exactly when it equals the current claim. Thus
\begin{equation}
    \label{eq:ci-sampler}
    \calS_\Phi(u;r,v)=
    \begin{cases}
        C_{\Next}(u,r), & (t,\alpha,v)=C_{\Now}(u)\\
        (t,\alpha,v), & (t,\alpha,v)\neq C_{\Now}(u)
    \end{cases}.
\end{equation}
For every fixed input and every $r$, replacement occurs with probability $1/q$. Consequently, for any input distribution and output event $E$,
\begin{equation}\label{eq:transition-action}
    \Pr{u'\in E}=\Pr{(t,\alpha,v)\in E}+\frac{\Pr{C_{\Next}(u,r)\in E}-\Pr{C_{\Now}(u)\in E}}{q}.
\end{equation}
Writing $\ell=\log{(T+1)}$, the circuit parameters are $m_\In=\ell+(m'+3)\log{q}$, $m_+=2\log{q}$, and $m_\Out=\ell+(m'+1)\log{q}$. 
Every bit string of the indicated length encodes a valid input or output. By \Cref{lemma:efficient-TQBF-arithmetization}, the field arithmetic, the round operations $\calO_t$ with their coordinates $j_t$, and the Boolean circuit for $P_T$ can all be constructed in polynomial time. These suffice to implement $\calS_\Phi$.

\vspace{1em}
Next, we analyze the behavior of our sampler $\calS_\Phi$ when the prover in \Cref{protocol:sum-check} is honest (\Cref{lemma:honest-prover-output-distribution}) or when the quantified Boolean formula $\Phi$ is false (\Cref{lemma:CIUniform-far}).
\begin{lemma}[Output distribution for the honest prover's replies]
    \label{lemma:honest-prover-output-distribution}
    Let $p$ be the probability distribution of $u=(t,\alpha,g)$ obtained as follows: choose $(t,\alpha)$ uniformly at random from $\cbra{0,\ldots,T}\times\bbF_q^{m'}$ and set $g \coloneqq \begin{cases}
    g_\star, & t<T\\
    0, & t=T
    \end{cases}$. Then, the output distribution $M_\calS p$ satisfies the following:
    \begin{itemize}
        \item If $\Phi$ is true, then $M_\calS p = U_{m_\Out}$. 
        \item If $\Phi$ is false, then $\TV(M_\calS p,U_{m_\Out}) = 1/(q(T+1))$. 
    \end{itemize}
\end{lemma}

\begin{proof}
    The initial tuple $(t,\alpha,v)$ is chosen uniformly at random. The current claim $C_{\Now}(u)$ has a pair $(t,\alpha)$ chosen uniformly at random and the correct claimed value $P_t(\alpha)$. The round of $C_{\Next}(u,r)$ moves forward cyclically, and its evaluation point is either unchanged or has one coordinate replaced by an independent field element chosen uniformly at random. Consequently, the first two coordinates of $C_{\Next}(u,r)$ are jointly uniform on $\cbra{0,\ldots,T}\times\bbF_q^{m'}$. For $t<T$, the next-round claim is correct because $g=g_\star$. For $t=T$, its claimed value in round $0$ is $1$.

    If $\Phi$ is true, $P_0=1$, so $C_{\Now}(u)$ and $C_{\Next}(u,r)$ have the same distribution. By \Cref{eq:transition-action} the sampler's output is therefore uniform. If $\Phi$ is false, $P_0=0$, and the only difference between the distributions of $C_{\Now}(u)$ and $C_{\Next}(u,r)$ occurs at round $0$, which has probability $1/(T+1)$ in both distributions. The evaluation point there has the same uniform distribution, but the claimed value changes from $0$ to $1$. As replacement has probability $1/q$, the sampler moves exactly $1/(q(T+1))$ of the total probability from the corresponding outputs with value $0$ to those with value $1$, which gives the claimed distance, as desired. 
\end{proof}

\begin{lemma}[Separation from uniform]
    \label{lemma:CIUniform-far}
    If $\Phi$ is false, every input distribution $p$ satisfies \[ \TV(M_\calS p,U_{m_\Out})\geq \frac{1}{4q(T+1)}. \]
\end{lemma}

\begin{proof}
Fix an arbitrary input distribution $p$. Draw $u=(t,\alpha,g)$ from $p$ and let $(t',\alpha',\nu)$ be the sampler's output. We bound the differences for the following two events:\footnote{Any convex combination of the two absolute differences, in particular the one with weights $1/(q+1)$ and $q/(q+1)$, gives a lower bound on the total variation distance for every distribution $p$.}
\begin{enumerate}[label={\upshape(\roman*)}]
    \setlength{\itemsep}{0pt}
    \item \label{thmitem:TQBF-to-CIUniform-soundness-i} Round $T$ is chosen ($t'=T$), which occurs with probability $1/(T+1)$ when $(t',\alpha',\nu)$ is drawn uniformly at random from $\cbra{0,\ldots,T}\times\bbF_q^{m'}\times\bbF_q$;
    \item \label{thmitem:TQBF-to-CIUniform-soundness-ii} $P_{t'}(\alpha')=\nu$, which occurs with probability $1/q$, for each fixed $t'$ and $\alpha'$. 
\end{enumerate}

Events \ref{thmitem:TQBF-to-CIUniform-soundness-i} and \ref{thmitem:TQBF-to-CIUniform-soundness-ii} give the following two absolute differences in event probabilities. Since each is at most $\TV(M_\calS p,U_{m_\Out})$, any weighted average of them with weights summing to one is also at most $\TV(M_\calS p,U_{m_\Out})$. We choose the weights $1/(q+1)$ and $q/(q+1)$, respectively, for a reason that will become clear below, and claim the following bound: 
\begin{equation}
    \label{eq:TQBF-to-CIUniform-no}
    \TV(M_\calS p,U_{m_\Out}) \geq \frac{1}{q+1}\abs*{ \frac{1}{T+1} - \Pr{t'=T} } + \frac{q}{q+1} \abs*{\frac{1}{q} - \Pr{P_{t'}(\alpha')=\nu}}.
\end{equation}

For event \ref{thmitem:TQBF-to-CIUniform-soundness-i}, namely $t'=T$, the sampler either stays at $t=T$ with probability $1-1/q$, or replaces the current claim at $t=T-1$ with probability $1/q$. Then, it follows that 
\begin{equation}
    \label{eq:CIUniform-far-event-i-bound}
    \Pr{t'=T}= \rbra*{1-\frac{1}{q}} \Pr{t=T} + \frac{1}{q}\,\Pr{t=T-1} \leq \Pr{t=T} + \frac{1}{q}.
\end{equation}
Under $U_{m_\Out}$, the round is $T$ with probability $1/(T+1)$, so \Cref{eq:CIUniform-far-event-i-bound} yields 
\begin{equation}
    \label{eq:CIUniform-far-event-i}
    \abs*{\frac{1}{T+1}-\Pr{t'=T}}\geq\frac{1}{T+1}-\Pr{t=T}-\frac{1}{q}.
\end{equation}

Now consider event \ref{thmitem:TQBF-to-CIUniform-soundness-ii}, namely $P_{t'}(\alpha')=\nu$. If $t=T$, the sampler either keeps $v\neq P_T(\alpha)$ or returns to round $0$ with the claim $P_0(\alpha)=1$. Both claims are false and therefore do not contribute to this event. 
We only need to consider the following two cases: (a) for each input with $t<T$, the initial value $v$, chosen uniformly at random, satisfies $v=P_t(\alpha)$ with probability $1/q$; and (b) replacement can increase this probability only when it removes a false current claim $v\neq P_t(\alpha)$ and produces a correct next-round claim $P_{t'}(\alpha')=\nu$. If the current claim is false, replacement occurs with probability $1/q$ and the next-round claim is correct with conditional probability at most $2/q$ by \Cref{eq:false-successor}, since $r$ is independent of $v$. Combining these two cases, we obtain
\begin{subequations}
\label{eq:CIUniform-far-event-ii-bound}
\begin{align}
    \Pr{P_{t'}(\alpha')=\nu}
    &\leq \Pr{t<T,\,v=P_t(\alpha)} + \Pr{t<T,\,v\neq P_t(\alpha),\,P_{t'}(\alpha')=\nu}\\
    &\leq \underbrace{\Pr{t<T} \cdot \frac{1}{q}}_{\text{(a) initial claim correct}} + \underbrace{\Pr{t<T} \cdot \frac{1}{q}\cdot\frac{2}{q}}_{\text{(b) false claim replaced by a correct claim}}\\
    &\leq \frac{1-\Pr{t=T}}{q}+\frac{2}{q^2}.
\end{align}
\end{subequations}

Under $U_{m_\Out}$, the value is correct with probability $1/q$, since exactly one value is correct for each round and evaluation point. We thus obtain that
\begin{equation}
    \label{eq:CIUniform-far-event-ii}
    \abs*{\frac1q-\Pr{P_{t'}(\alpha')=\nu}}\geq\frac{\Pr{t=T}}q-\frac{2}{q^2}. 
\end{equation}

Substituting the bounds for events \ref{thmitem:TQBF-to-CIUniform-soundness-i} and \ref{thmitem:TQBF-to-CIUniform-soundness-ii} from \Cref{eq:CIUniform-far-event-i,eq:CIUniform-far-event-ii} into \Cref{eq:TQBF-to-CIUniform-no}, the chosen weights $1/(q+1)$ and $q/(q+1)$ cancel the unknown term $\Pr{t=T}$. Since $q=8(T+1)$ by \Cref{lemma:efficient-TQBF-arithmetization}, we obtain
\[
 \TV(M_\calS p,U_{m_\Out})
 \geq\frac{1}{q+1}\left(\frac{1}{T+1}-\frac{3}{q}\right)
 =\frac{5}{8(q+1)(T+1)}
 \geq\frac{1}{4q(T+1)} = \frac{1}{32(T+1)^2}. \qedhere
\]
\end{proof}

\vspace{1em}
\begin{proof}[Proof of \Cref{thm:CIUniform-PSPACE-hard}]
Since \TQBF{} is \PSPACE{}-complete~\cite[Theorem~4.3]{SM73}, it suffices to give a Karp reduction from \TQBF{} to \CIUniform{}. Given a \TQBF{} instance $\Phi$, output the Boolean circuit for the sampler $\calS_\Phi$ defined in \Cref{eq:ci-sampler}. By \Cref{lemma:efficient-TQBF-arithmetization}, the field arithmetic, the round operations, and the circuit for $P_T$ used by $\calS_\Phi$ can be constructed in time polynomial in $|\Phi|$.

Let $n=m_\In+m_\Out$ as in \Cref{def:CIUniform}. The output length $m_\Out=\log{(T+1)}+(m'+1)\log{q}$ gives $n\geq m'+1$. The choice of $T$ in the proof of \Cref{lemma:efficient-TQBF-arithmetization} therefore gives $T+1\leq m'(m'+1)\leq n^2$.

If $\Phi$ is true, \Cref{lemma:honest-prover-output-distribution} gives an input distribution $p$ with $M_{\calS_\Phi}p=U_{m_\Out}$. If $\Phi$ is false, \Cref{lemma:CIUniform-far} and $q=8(T+1)$ give, for every input distribution $p$,
\[
    \TV(M_{\calS_\Phi}p,U_{m_\Out})
    \geq \frac{1}{4q(T+1)}
    = \frac{1}{32(T+1)^2}
    \geq \frac{1}{32n^4}.
\]
Hence $\Phi\mapsto\calS_\Phi$ maps true and false formulas to \emph{yes} and \emph{no} instances of $\CIUniform[1/(32n^4)]$, respectively.
\end{proof}


\section{\CIUniform{} is in \qqStoqAM{}}\label{sec:containment}

We begin by defining the stoquasti version of \qqQAM{}: 

\begin{definition}[\qqStoqAM{}]
\label{def:qq-StoqAM}
Let $c(n)$ and $s(n)$ be functions of the input length $n \coloneqq |x|$ such that $0 \leq s(n) < c(n) \leq 1$. Let $\ell(n)$ be a positive integer-valued function such that $\ell(n) \leq \poly(n)$.
A promise problem $\calI = (\calI_{\yes}, \calI_{\no})$ is in $\qqStoqAM\sbra*{\ell,c,s}$, if there exists an efficient quantum verifier $V(x)=(V_1(x),V_2(x))$ and a polynomial $p$, where $V_1$ prepares $p(n)$ EPR pairs and sends their second halves to the prover, and stores their first halves in $\sfV$, and $V_2$ is a stoquastic verification circuit as in \Cref{def:stoquastic-verification} acting on the message register $\sfM$ and $\sfV$, such that:
\begin{itemize}
    \item \textbf{\emph{Completeness}}. For every $x \in \calI_{\yes}$, there exists a prover $P(x)$ such that, upon receiving the halves of the EPR pairs, $P(x)$ returns an $\ell(n)$-qubit state and
    \[\Pr{(\protocol{P}{V})(x)\text{ accepts}} \geq c(n).\]
    \item \textbf{\emph{Soundness}}. For any $x \in \calI_{\no}$ and any prover $P(x)$ that returns an $\ell(n)$-qubit state,
    \[\Pr{(\protocol{P}{V})(x)\text{ accepts}} \leq s(n).\]
\end{itemize}
\end{definition}

As in \Cref{def:qqQAM}, we omit the reply length $\ell$ and write
\[ \qqStoqAM[c,s] \coloneqq \cup_{\ell(n) \leq \poly(n)} \qqStoqAM[\ell,c,s]. \]

To adopt the convention $\qqStoqAM \coloneqq \qqStoqAM[1,3/4]$, we need parallel repetition for \qqStoqAM{}, with the proof deferred in \Cref{sec:qqStoqMA-parallel-repetition}: 
\begin{restatable}[Parallel repetition for \qqStoqAM{}]{lemma}{qqStoqAMParallelRepetition}
    \label{lemma:qqStoqAM-parallel-repetition}
    Let $c(n)$ and $s(n)$ be threshold functions such that $1/2\leq s(n)<c(n)\leq 1$, and let $r(n) \geq 1$ be an efficiently computable function that is positive integer-valued and polynomially bounded. Then
    \[ \qqStoqAM[c,s]\subseteq\qqStoqAM\sbra*{\frac{1+c^r}{2},\frac{1+s^r}{2}}. \]
\end{restatable}

\begin{remark}[Nonnegative joint states can be suboptimal]
Consider a verifier that prepares an EPR pair on $(\sfB,\sfB'')$, sends $\sfB''$ to the prover, and receives a qubit $\sfA$. It swaps $\sfB$ with a fresh $\ket{0}$ controlled on $\sfA=1$, then measures $\sfB$ in the Hadamard basis and accepts on $+$. For the joint state $\rho$ on $(\sfB,\sfA)$, let $\sigma_b=(I\otimes\bra{b})\rho(I\otimes\ket{b})$. The marginal condition gives $\sigma_0+\sigma_1=I/2$, so the acceptance probability is $\Tr(\ketbra{+}{+}\sigma_0)+\tfrac12\Tr(\sigma_1)=\tfrac12+\tfrac12\Tr(X\sigma_0)\leq\tfrac34$. 
Here $\Tr(X\sigma_0)=\braket{+}{\sigma_0}{+}-\braket{-}{\sigma_0}{-}\leq1/2$. Applying a Hadamard to $\sfB''$ and returning it gives $\sigma_0=\ketbra{+}{+}/2$ and attains $3/4$. If $\rho$ is entrywise nonnegative in the computational basis, the off-diagonal entries of $\sigma_0$ and $\sigma_1$ are nonnegative and sum to zero. Then $\Tr(X\sigma_0)=0$, giving acceptance probability $1/2$.
\end{remark}

\vspace{1em}
For $\CIUniform[\eta]$, we index both the promise and the verifier's soundness by $n=m_\In+m_\Out$ as in \Cref{def:CIUniform}. We take $\eta(n)>0$ so that the two cases of the promise are disjoint.

\begin{theorem}\label{thm:ci-stoq}
    $\CIUniform[\eta]\in\qqStoqAM[1,1-\eta^2/2]$.
\end{theorem}

\begin{proof}[Proof of \Cref{thm:main-informal}]
By \Cref{thm:CIUniform-PSPACE-hard,thm:ci-stoq}, the reduction from $\TQBF$ gives a perfectly complete \qqStoqAM{} verifier with soundness at most $1-\eta(n)^2/2$, where $\eta(n)=1/(32n^4)$. Taking $r=2/\eta(n)^2$ copies in \Cref{lemma:qqStoqAM-parallel-repetition} preserves perfect completeness and gives soundness at most $(1+(1-1/r)^r)/2<3/4$. The sampler width $n$ is polynomially bounded in the formula length, so this gives $\PSPACE\subseteq\qqStoqAM_1$.

For a $\qqStoqAM_1$ verifier with soundness $s$, a $\qqQAM$ verifier can run polynomially many copies in parallel and accept if all copies accept. By \Cref{eq:stoq-all-copies-accept}, $k$ copies have soundness at most $s^k\leq 1/2$ for $k\geq 1/(1-s)$ and preserve perfect completeness. Thus, using $\QIP=\PSPACE$~\cite{JJUW11}, we obtain
\[
    \PSPACE\subseteq\qqStoqAM_1\subseteq\qqQAM\subseteq\QIPtwo\subseteq\QIP=\PSPACE.\qedhere
\]
\end{proof}

\subsection{A \qqStoqAM{} proof system for \CIUniform{}}

We specialize the image test of~\cite[Lemma~4.1]{KLGN19} to the sampler $\calS$ and implement the verifier's final measurement by a SWAP test. Let $U_\calS$ be the classical reversible implementation of the sampler that acts on the input, seed, and output registers $(\sfA,\sfR,\sfB)$ by
\[
U_\calS\ket{u,r,w}=\ket{u,r,w\oplus\calS(u,r)}.
\]
The verifier computes $\calS(u,r)$ into fresh zero qubits, adds the result to $\sfB$ by bitwise XOR, and reverses the computation on the fresh qubits. This gives $U_\calS=U_\calS^\dagger$ using polynomially many Toffoli, CNOT, and X gates.

\LinesNotNumbered
\begin{algorithm}[ht!]
  \caption{A two-message stoquastic verifier for $\CIUniform[\eta]$}\label[algocf]{prot:ci}
  \SetEndCharOfAlgoLine{.}
  \textbf{1.} Prepare $m_\Out$ EPR pairs across the registers $(\sfB,\sfB'')$ and send $\sfB''$ to the prover\;
  \textbf{2.} Receive the $m_\In$-qubit input register $\sfA$ and the $m_+$-qubit seed register $\sfR$. Apply $U_\calS^\dagger$ to the registers $(\sfA,\sfR,\sfB)$\;
  \textbf{3.} Apply the SWAP test between $(\sfR,\sfB)$ and fresh reference registers $(\sfR',\sfB')$ initialized to $\ket{+}^{\otimes m_+}\ket{0}^{\otimes m_\Out}$, using a control qubit $\sfO$ initialized to $\ket{+}$. Accept if and only if measuring $\sfO$ in the Hadamard basis gives outcome $+$\;
\end{algorithm}

The verifier prepares each EPR pair by applying a CNOT to $\ket{+}\ket{0}$. Together with \Cref{lemma:swap-test}, the reversible implementation of $U_\calS$ gives a polynomial-time uniform stoquastic verifier as in \Cref{def:qq-StoqAM}. The verifier can pad both messages to fixed polynomial lengths and ignore the extra registers.

\subsection[Analysis of Protocol~\ref*{prot:ci}]{Analysis of \Cref{prot:ci}}\label{sec:protocol-analysis}

\begin{proof}[Proof of \Cref{thm:ci-stoq}]
Let $\rho$ be the joint state on $(\sfA,\sfR,\sfB)$ immediately before the verifier applies $U_\calS^\dagger$ in \Cref{prot:ci}.

\parheading{Acceptance probability.}
Write $\ket{\omega}=\ket{+}^{\otimes m_+}\ket{0}^{\otimes m_\Out}$ for the reference state and $\Pi=U_\calS(I_{\sfA}\otimes\ketbra{\omega}{\omega})U_\calS^\dagger$ for the projector onto the subspace tested by the verifier. After applying $U_\calS^\dagger$, the state on $(\sfR,\sfB)$ is $\sigma=\Tr_{\sfA}(U_\calS^\dagger\rho U_\calS)$. By the SWAP test in \Cref{lemma:swap-test}, the acceptance probability is
\begin{align}
    \label{eq:ci-acceptance-probability}
    p_{\rm acc} = \frac12\bigl(1+\braket{\omega}{\sigma}{\omega}\bigr) 
    = \frac12\bigl(1+\Tr[(I_{\sfA}\otimes\ketbra{\omega}{\omega})U_\calS^\dagger\rho U_\calS]\bigr) 
    = \frac12\bigl(1+\Tr(\Pi\rho)\bigr). 
\end{align}

\parheading{Perfect completeness.}
Suppose $M_\calS p=U_{m_\Out}$ for some input distribution $p$. Upon receiving $\sfB''$, the prover maps each $\ket{u'}$ to the following state on $(\sfA,\sfR)$ and returns these registers:
\[
    \ket{k_{u'}}=2^{(m_\Out-m_+)/2}\sum_{(u,r):\,\calS(u,r)=u'}\sqrt{p(u)}\,\ket{u}_{\sfA}\ket{r}_{\sfR}.
\]
For each $u'$, we have $\norm{\ket{k_{u'}}}^2=2^{m_\Out}(M_\calS p)(u')=1$. Distinct outputs have disjoint sets of preimages, so the vectors $\ket{k_{u'}}$ are orthonormal and the map is an isometry. Applying $U_\calS^\dagger$ to the resulting joint state gives
\begin{align*}
    U_\calS^\dagger\left(2^{-m_\Out/2}\sum_{u'}\ket{k_{u'}}_{\sfA\sfR}\ket{u'}_{\sfB}\right)
    &= 2^{-m_+/2}\sum_{u,r}\sqrt{p(u)}\,\ket{u}_{\sfA}\ket{r}_{\sfR}\ket{0}^{\otimes m_\Out}_{\sfB}\\
    &= \left(\sum_u\sqrt{p(u)}\ket{u}_{\sfA}\right)\ket{\omega}_{\sfR\sfB}.
\end{align*}
The SWAP test compares two copies of $\ket{\omega}$ and accepts with certainty.

\parheading{Soundness.}
Suppose $\TV(M_\calS p,U_{m_\Out})\geq\eta$ for every input distribution $p$. Fix any prover strategy and purify $\rho$ to $\ket{\psi}$ using a private register $\sfQ$. The prover acts only on $\sfB''$ and private registers, so the verifier's register $\sfB$ is still maximally mixed. Measuring $\sfB$ in the computational basis on $\ket{\psi}$ gives $U_{m_\Out}$.

To bound the projection of $\ket{\psi}$ onto the image of $\Pi\otimes I_{\sfQ}$, take any unit vector $\ket{\xi}$ in that image. By the definition of $\Pi$, applying $U_\calS^\dagger$ to $\ket{\xi}$ gives $\ket{\omega}$ on $(\sfR,\sfB)$ and a unit vector $\sum_u\ket{u}_{\sfA}\ket{\chi_u}_{\sfQ}$ on $(\sfA,\sfQ)$. Thus $p(u)=\norm{\ket{\chi_u}}^2$ defines an input distribution. Applying $U_\calS$ again writes $\calS(u,r)$ into $\sfB$ for each pair $(u,r)$. Since distinct pairs are orthogonal in $(\sfA,\sfR)$, measuring $\sfB$ on $\ket{\xi}$ gives $u'$ with probability
\[
    2^{-m_+}\sum_{(u,r):\,\calS(u,r)=u'}p(u)=(M_\calS p)(u').
\]
Measurement cannot increase trace distance. Comparing these two distributions and using the trace distance formula for pure states gives
\begin{align*}
    \eta \leq \TV(M_\calS p,U_{m_\Out})
    \leq \frac12\norm*{\ketbra{\psi}{\psi}-\ketbra{\xi}{\xi}}_1
    = \sqrt{1-\abs{\innerprod{\psi}{\xi}}^2}.
\end{align*}
Hence $\abs{\innerprod{\psi}{\xi}}^2\leq 1-\eta^2$ for every such $\ket{\xi}$. The maximum squared overlap with a unit vector in a subspace equals the squared norm of the orthogonal projection onto that subspace. Substituting into \Cref{eq:ci-acceptance-probability} gives
\[
    p_{\rm acc}=\frac12\bigl(1+\norm*{(\Pi\otimes I_{\sfQ})\ket{\psi}}^2\bigr)\leq 1-\frac12\eta^2.\qedhere
\]
\end{proof}

\section*{Acknowledgments}
\noindent
The author was supported in part by funding from the Swiss State Secretariat for Education, Research and Innovation (SERI). 
This work was also supported in part by a grant of access to OpenAI models through the ChatGPT for Academic Researchers program.

\section*{AI use disclosure}
\noindent
Large language model tools were used interactively at all stages of the preparation of this manuscript: from the earliest exploratory research stages to the final exposition and writing steps. The final manuscript was substantially revised by the authors, who remain solely responsible for all mathematical claims, proofs, references, and conclusions.

\bibliographystyle{alphaurlQ}
\bibliography{qqStoqAM}

\appendix
\section{Parallel repetition for two-message stoquastic proof systems}
\label{sec:qqStoqMA-parallel-repetition}

\qqStoqAMParallelRepetition*

\begin{proof}
Fix a $\qqStoqAM[c,s]$ proof system and an input $x$ of length $n$. Write $\omega(x)$ for the maximum acceptance probability on $x$ and let $r=r(n)$. The verifier prepares $r$ independent copies of the EPR challenge, sends all second halves in one message, and runs the reversible gates of each copy on the joint reply and the stored EPR halves. It then performs a single SWAP test between the $r$ unmeasured output qubits and a fresh register in $\ket{+}^{\otimes r}$, as in~\cite[Theorem~3.14 and Figure~2]{LW26}. By \Cref{lemma:swap-test}, this last step uses only Toffoli and CNOT gates and one Hadamard-basis measurement. The repeated verifier is polynomial-time uniform because $r$ is polynomially bounded and efficiently computable.

For any joint prover strategy $P$, let $\sigma_P$ be the state of the $r$ output qubits just before the SWAP test. If the verifier measured each output in the Hadamard basis, the probability of obtaining $+$ in every copy would be $a(P)=\Tr[\sigma_P(\ketbra{+}{+})^{\otimes r}]$. The SWAP test therefore accepts with probability $(1+a(P))/2$ by \Cref{lemma:swap-test}.

The individual measurements are independent verifier tests, while $P$ may act jointly across all $r$ copies. Applying the product rule for the maximum probability of prescribed outcomes~\cite[Section~4(a)]{MW12} to the outcome $+$ in every copy gives
\begin{equation}\label{eq:stoq-all-copies-accept}
    \max_P a(P)=\omega(x)^r.
\end{equation}
The maximum acceptance probability of the repeated verifier is therefore $(1+\omega(x)^r)/2$. On a \emph{yes} instance, $\omega(x)\geq c(n)$ gives completeness at least $(1+c(n)^r)/2$. On a \emph{no} instance, $\omega(x)\leq s(n)$ gives soundness at most $(1+s(n)^r)/2$.
\end{proof}

\end{document}